\documentclass[prb,twocolumn,showpacs,preprintnumbers, letter]{revtex4}

\usepackage{graphicx} 
\usepackage{dcolumn} 

\begin{document}
\title{Asymmetry in self-assembled quantum dot-molecules
made of identical InAs/GaAs quantum dots}

\author{Lixin He}
\author{Gabriel Bester}
\author{Alex Zunger}
\affiliation{National Renewable Energy Laboratory, Golden, Colorado 80401}

\date{\today}
\begin{abstract}

We show that a diatomic dot molecule made of two
identical, vertically stacked, strained InAs/GaAs
self-assembled dots exhibits 
an asymmetry in its single-particle and may-particle
wavefunctions.
The single-particle wave function is asymmetric
due to the inhomogeneous strain, while the asymmetry of 
the many-particle wavefunctions is caused by the
correlation induced localization:
the lowest singlet $^1\Sigma_g$ and
triplet $^3\Sigma$ states show  
that the two electrons are each localized on
different dots within the molecule, 
for the next singlet states $^1\Sigma_u$ 
both electrons are localized on the same (bottom) dot for interdot
separation $d>$ 8 nm. 
The singlet-triplet splitting is found to be $\sim 0.1$ meV at inter-dot
separation $d$=9 nm and 
as large as 100 meV for $d$=4 nm, 
orders of magnitude larger than the few meV found in the large 
(50 - 100 nm) electrostatically confined dots.

\end{abstract}



\maketitle 

Quantum dot-molecules (QDM) occupied by two electronic spins 
have been proposed as 
a basis for quantum
computation.\cite{divincenzo95} 
Loss and DiVincenzo~\cite{loss98} proposed a ``swap gate'' based on 
a simplified model in which the two localized spins have 
Heisenberg coupling 
$H = J_{S-T} \vec{S}_1 \cdot \vec{S}_2$,
where $\vec{S}_1$ and $\vec{S}_2$ are the spin-${1\over2}$ operators
for the two localized electrons, 
and $J_{S-T}$ is the effective Heisenberg exchange
splitting, being 
the difference in energy between the spin-triplet state
with total spin $S=1$ and the spin-singlet with $S=0$.
Successful operation would require a large-singlet triplet splitting
$J_{S-T}$ (fast swap time~\cite{loss98}), 
and that the probability 
$Q_{\rm tot}^{(\nu)}$ of the two electrons in state $\nu$
simultaneously occupying one dot be small (maximizing
entanglement~\cite{schliemann01}).
The search for a nanosystem with large
$J_{S-T}$ and small $Q_{\rm tot}^{(\nu)}$ involves engineering of
the properties of
the corresponding many-particle wave functions.
In a simplified molecular model, the 
single-particle wave functions of the two electrons 
are given by bonding
$\psi_{\sigma_g}=(\chi_T+\chi_B)/\sqrt{2}$ and antibonding 
$\psi_{\sigma_u}=(\chi_T-\chi_B)/\sqrt{2}$ molecular orbitals,
constructed from the individual orbitals $\chi$
of the top (T) and bottom (B) dot.
The many-particle states are then the corresponding
product states
$|\sigma_g^{\uparrow}\sigma_g^{\downarrow}\rangle\sim$ $^1\Sigma_g$
(singlet), and $|\sigma_g^{\uparrow}\sigma_u^{\uparrow}\rangle\sim$ 
$^3\Sigma$ (triplet), etc. 
In this picture both single-particle molecular orbitals and 
the many-particle states are delocalized on both dots constituting the
dot-molecule.
Here, we discuss via atomistic single-particle and many-body calculations
two important deviations from this simplified molecular picture, leading to 
asymmetries both in the
single-particle molecular orbitals due to the inhomogeneous strains,
and in the many-body states (i.e. localization either on T or on B)
as a consequence of correlation.
For the many-body states 
we find Mott-like transitions for the 
first and third singlet states: both electrons are localized 
on one dot at large $d$ and delocalized over both dots are small $d$.
The double occupancy $Q_{\rm tot}^{(\nu)}$ of the first singlet state 
is surprisingly large ($\simeq$40\%) for an interdot separation of 5~nm.
The triplet states and the second singlet state are Mott-localized 
at every interdot separations with $Q_{\rm tot}^{(\nu)}$=0 (no double
occupation) for the triplet states and $Q_{\rm tot}^{(\nu)}$=1 
for the singlet state.

\begin{figure}
\includegraphics[width=3.5in,angle=0]{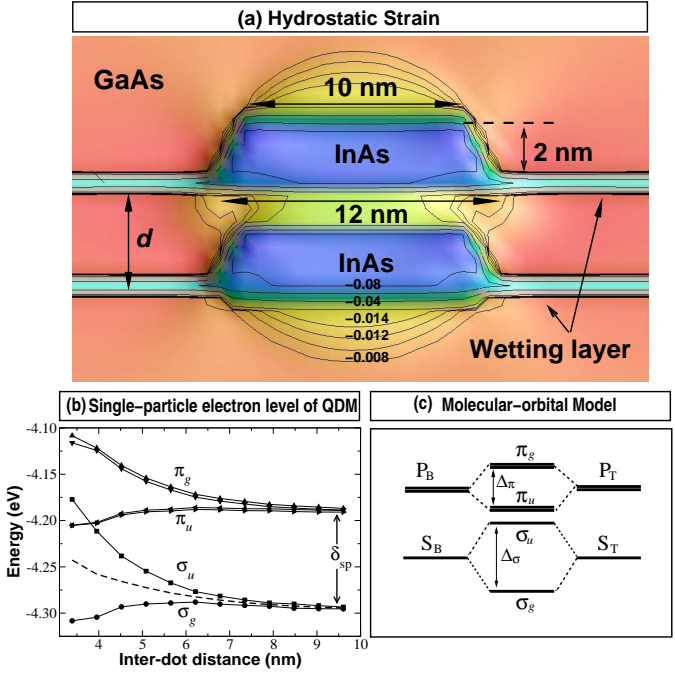}
\caption{
(a) (Color online)
Contour plot of the hydrostatic strain ${\rm Tr} (\epsilon)$ in the 
two vertically coupled quantum dots.
The inter-dot distance $d$ is measured from one 
wetting layer to the next. The iso-strain values are also
marked in the figure.
(b) Molecular orbital energy levels vs. inter-dot distance.
The dashed line is the average of $\sigma_g$ and $\sigma_u$.
$\delta_{sp}$ is the single dot s-p energy level splitting.
(c) Sketch of bonding-antibonding splitting. 
$S_T$, $S_B$ are ``s'' while $P_T$ and $P_B$ are 
``p'' single dot orbitals on top and
bottom dots. 
Here $S_T+S_B=\sigma_g$ and $P_T+P_B=\pi_u$ are bonding states, while
the $S_T-S_B=\sigma_u$ and $P_T-P_B=\pi_g$ are antibonding states. 
}
\label{fig:geom}
\end{figure}

Previous models of dot-molecules have focused on 
electrostatically-confined dots\cite{ashoori92,johnson92,tarucha96},
which have a very large confining dimension of 50 - 100 nm. 
Such dots exhibit typical 
single-particle level spacings of $\Delta \epsilon_{e}$=3 - 5 meV, 
Coulomb energies $J_{ee}$ of about 5
meV $\geq$ $\Delta \epsilon_{e}$, whereas exchange
energies and correlation energies are around 1 meV. 
Many experiments
were recently done on {\it two} such coupled dots\cite{pi01,rontani04}, showing
that the splitting between the bonding and anti-bonding molecular levels is as
large as 3.5 meV at an interdot separation of 2.5 nm, comparable to the
single-particle energy spacing
$\Delta \epsilon_{e}$ of the single dot. 
Because of the very large size of such dots, their single-particle levels can
be described by simple one-band effective mass ``particle in a box'' 
models using
the external potential generated by a combination of 
band offset, the gate potential and the
ionized impurities.\cite{fonseca98,bednarek03} 
Alternatively, one can simply assume a
particle-in-a-parabolic well model.\cite{hu00,rontani01} 
In these descriptions, 
the single-particle states are symmetric , but many-electron
symmetry breaking is possible due to correlation effects, as shown via
unrestricted Hartree-Fock  treatment of the effective-mass approximation
(UHF-EMA),
\cite{yannouleas99}, configuration-interaction treatment of the
effective-mass approximation (CI-EMA), \cite{rontani01} 
or Mott-Hubbard model. \cite{stafford94} 

Here, we discuss localization
effects and singlet-triplet splitting of electrons in 
vertically coupled self-assembled InAs/GaAs quantum dot
molecules grown epitaxially\cite{bimberg_book,ota03,ota04a}.
Such dots have much smaller confining dimensions (height of 3 - 5 nm), and when
made of InAs/GaAs their electronic level splitting is 
$\Delta \epsilon_{e} \sim$ 40 - 50 meV,  {\it larger} than the 
interelectronic Coulomb repulsion $J_{ee} \sim$ 10
-20 meV, and the exchange energy $K_{ee} \sim$ 2 - 5 meV. 
In this paper, we show that in
vertically aligned self-assembled dots, 
one can achieve singlet-triplet
splittings of up to 100 meV.

Figure~\ref{fig:geom}a shows the geometry selected for the dot molecules,
consisting of two-dimensional InAs wetting layers, a pair of
2 nm tall InAs dots in the shape of truncated cones embedded in 
a GaAs matrix.
The considered dots are identical and describe the
experimental extreme case of perfect growth. In reality, both dots can
be geometrically and compositionally different. This extrinsic 
inequivalence would
reinforce the intrinsic asymmetry of the system which our results describe.
The hydrostatic strain field ${\rm Tr} (\epsilon)$, with the iso-strain
lines shown in Fig.~\ref{fig:geom}a, 
is calculated atomistically by relaxing
the bond-stretching and bond-bending forces according to the valence force
field model (VFF).\cite{keating66,martins84} 
It is clearly seen in Fig.~\ref{fig:geom}a that both dots have large and
nearly constant hydrostatic strain inside the dots which decays rapidly 
outside the dots. 
However, even though the dots comprising the molecule are the same,
the strain on the two dots is different 
since the molecule lacks inversion symmetry. We see that
the top dot is slightly more strained than the bottom dot. 
Furthermore, the GaAs region {\it between} the two dots is
much more strained than in other parts of the matrix.

Having established a realistic geometry and the
relaxed atomic positions $\{{\bf R}_{m,\alpha}\}$, 
we calculate the
single-particle electronic structure by constructing a pseudopotential  
$V({\bf r}) =\sum_{m,\alpha} v_{\alpha}({\bf r}- {\bf R}_{m,\alpha})$ 
from a superposition of
screened atomic potentials $v_{\alpha}$ of species $\alpha$= Ga, In, As.
Here, $v_{\alpha}$ is constructed~\cite{williamson00} by fitting to available
experimental data the bulk InAs and GaAs band energies,
effective-masses, hydrostatic and biaxial deformation potentials, and
band-offsets. The pseudopotentials used in present work are taken from 
Ref.~\onlinecite{williamson00}.
The Schr\"{o}dinger equation is solved by the Linear Combination
of Bulk Bands (LCBB)~\cite{wang99b} method in a basis
$\{\phi_{n,{\bf k},\tensor{\epsilon}}^{(\lambda)}({\bf r})\}$ of 
Bloch orbitals, i.e.
$\psi_{i}({\bf r}) =\sum_{n,{\bf k},\lambda} C_{n,{\bf k},\lambda}^{(i)}\,
\phi_{n,{\bf k},\tensor{\epsilon}}^{(\lambda)}({\bf r})$
of band index $n$ and wave vector ${\bf k}$ of material $\lambda$ 
(= InAs, GaAs). The basis functions are strained
uniformly to constant but different strains $\tensor{\epsilon}$. 
We use  $\tensor{\epsilon}=0$ for the
(unstrained) GaAs matrix material, and an average $\tensor{\epsilon}$ value 
from VFF for the strained dot material (InAs).
For the InAs/GaAs system, we use $n=2$ for electron states on a 
6$\times$6$\times$28 k-mesh.
Unlike the effective-mass description of single-particle state used for
electrostatic dot,\cite{fonseca98,bednarek03,hu00,rontani01} 
here we allow inter-band and inter-valley
coupling.

\begin{figure}
\includegraphics[width=3.5in,angle=0]{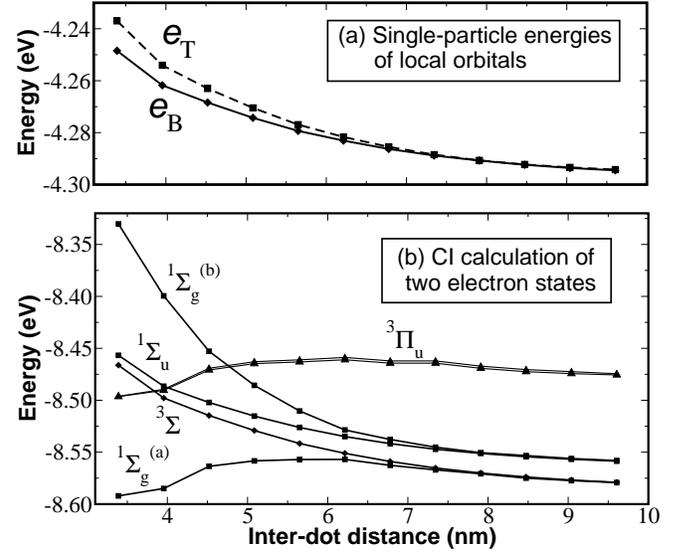}
\caption{
(a) The effective single-particle energy levels
$e_{\rm T}$ and  $e_{\rm B}$ of 
dot-centered orbitals on the top dot and bottom dot respectively. 
(b) Energy of two-electron states calculated from CI using 
all confined molecular orbitals.
}
\label{fig:e1-e2}
\end{figure}

{\it Asymmetry of single-particle states}:
Figure \ref{fig:geom}b shows the single-particle dot-molecule
energy levels as a function of inter-dot distance $d$.
These results can be generally understood using bonding/antibonding molecular
orbitals as shown in Fig.\ref{fig:geom}c. 
At large $d$, the energy levels converge to those of
single dots, showing an s-level and, at $\delta_{sp}=106$ meV higher, 
two p-levels 
split by a few meV reflecting the atomistic $C_{2v}$ symmetry of the
cylindrically symmetric dots\cite{bester05a}. 
Bonding and antibonding molecular orbitals form when the two dots interact:
the two single-dot s orbitals form
molecular $\sigma_g$ and $\sigma_u$ orbitals, 
whereas the four single-dot p orbitals (two on each dot)
form two molecular $\pi_u$ and two molecular 
$\pi_g$ orbitals.\cite{rontani01} 
Note that the splitting of $\sigma_g$ from $\sigma_u$ is not symmetric, as
can be seen by looking at the average of both energies  in
Fig.~\ref{fig:geom}b (dashed line). The average energy increases
with decreasing interdot separation  in response to the 
strain exerted by the presence of the other dot (Fig.\ref{fig:geom}a).
Beyond this overall effect, the strain field is different on 
both geometrically identical dots because the dots are non-spherical 
(Fig.\ref{fig:geom}a) and lack inversion symmetry. This
causes an asymmetry of the molecular orbitals at
short interdot distances. At $d\sim$ 3.4 nm, we found that
the bonding (anti-bonding) states are slightly more 
localized on the bottom (top) dot. 
In previous effective-mass calculations,
\cite{pi01,rontani04,hu00,rontani01,yannouleas99}  
strain effects were not included and single-particle asymmetry 
was not found.   
Particularly, in Ref.\onlinecite{pi01}, an {\it ad hoc} parameter 
was introduced to force the asymmetry of the dot molecule.

A more quantitative analysis of the asymmetry can be given when the
molecular orbitals are transformed via a Wannier-like transformation into
single dot states.
These can be obtained from a unitary rotation of {\it molecular} 
orbitals $\psi_{i}$, 
\begin{equation}
\chi_{l,p} =\sum_{i} \mathcal{U}_{l,p}^{(i)}\, \psi_{i} \, ,
\label{eq:wannier}
\end{equation}
where, $\psi_{i}$ is the $i$-the molecular orbital
and $\chi_{l,p}$ are the rotated, dot-centered orbitals
(the $l$-th orbital localized on $p$= T or B dot).
$\mathcal{U}$ are unitary matrices, $\mathcal{U}^{\dag}\mathcal{U}=I$,
chosen to maximize the total self-Coulomb energy.\cite{edmiston63,he05d} 
Once $\mathcal{U}_{l,p}^{(i)}$ 
are known, we define the orbital energies of
the dot-centered states $\chi_{l,p}$ as:
\begin{equation}
e_{l,p}=\langle \chi_{l,p} |\hat{T}|\chi_{l,p}\rangle=\sum_{i} 
(\mathcal{U}_{l,p}^{(i)})^*\;
\mathcal{U}_{l,p}^{(i)}\; \epsilon_{i} \; ,
\label{eq:sp-energy}
\end{equation}
where $\hat{T}$ is the kinetic energy operator, and
$\epsilon_{i}$ is the energy of the $i$-th molecular orbital. 
The energies $e_{l,p}$ of the dot-centered orbitals are
depicted in Fig.~\ref{fig:e1-e2}a 
as a function of the interdot separation. 
We see that the single-particle energies 
for both B and T orbitals rise quickly   
as the inter-dot distance is reduced,
but the energy of the top dot orbital
raises faster. 
At $d\sim$ 3.4 nm, there is an energy splitting of $\sim$ 12 meV
between top and bottom dot orbitals, which causes  
the asymmetry of the wave functions between top and
bottom dots. 
 
\begin{figure}
\includegraphics[width=3.5in,angle=0]{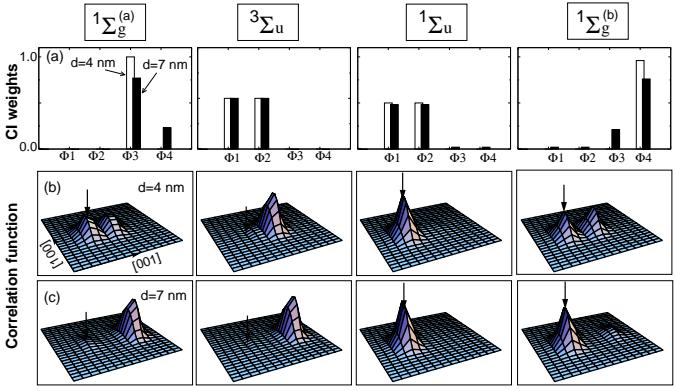}
\caption{(Color online) 
Panel (a) gives the weights of configurations $\Phi_1$, $\Phi_2$, 
$\Phi_3$ and $\Phi_4$  of the many-particle CI wave functions. 
Panels (b) and (c) depict the probability of finding the
second electron at position {\bf r} given that the first electron 
has been found on the center of the bottom
dot (indicated by the arrows)
for the interdot distances (b) d=4 nm and (c) d= 7 nm.  
}
\label{fig:corrfunc}
\end{figure}

{\it Many-particle symmetry-breaking}: 
Having obtained the ``molecular'',
single-particle energy (Fig.~\ref{fig:geom}b),
and wave functions, we calculate all interelectronic
Coulomb and exchange integrals $J$ and $K$ of $\psi_i$
by numerical
integration\cite{Franceschetti97}, and set up a screened
configuration-interaction 
expansion\cite{franceschetti99}.
A microscopic position-dependent dielectric screening\cite{franceschetti99} 
is applied to both Coulomb and exchange integrals
to represent the inner electrons that are not calculated explicity. 
Considering six molecular orbitals $\sigma_g$,
$\sigma_u$, $\pi_u$, $\pi_g$ of Fig.~\ref{fig:geom}b, we have a total of 66
Slater determinants. The many-body wave functions $\Psi_{\nu}$
are written as linear
combinations of these determinants $|\Phi_{\mathcal{C}}\rangle$ as,
%
$\Psi_{\nu} = \sum_{\mathcal{C}} A_{\nu}(\mathcal{C})\,
|\Phi_{\mathcal{C}}\rangle$.
%
The resulting many-particle energies are shown as a
function of interdot separation in Fig.~\ref{fig:e1-e2}b. 
The energy splittings  $J_{S-T}$  between the ground state 
singlet $^1\Sigma_g^{(a)}$  and triplet 
$^3\Sigma$ ranges from 0 - 100 meV and is much  larger 
than in electrostatic dot molecules
($<$ 1 meV).\cite{ashoori92,johnson92,tarucha96}
In Fig.\ref{fig:corrfunc}a we decompose the two-electron wave 
functions  into the leading configurations
$\Phi_1= |\sigma_g^{\uparrow}\sigma_u^{\downarrow}\rangle$,
$\Phi_2= |\sigma_g^{\downarrow}\sigma_u^{\uparrow}\rangle$,
$\Phi_3= |\sigma_g^{\uparrow}\sigma_g^{\downarrow}\rangle$,
and
$\Phi_4= |\sigma_u^{\uparrow}\sigma_u^{\downarrow}\rangle$. 
The ground state is the singlet $^1\Sigma_g^{(a)}$ state, 
followed by the 
three-fold degenerated triplet states $^3\Sigma$ 
(we depict only the $s_z$=0 state made of 
$\Phi_1+\Phi_2$ in Fig.\ref{fig:corrfunc})
and the next
singlets $^1\Sigma_u$ (made of $\Phi_1-\Phi_2$)
and $^1\Sigma_g^{(b)}$. 
To explore the localization of these states,
we plot in Fig.\ref{fig:corrfunc}b and \ref{fig:corrfunc}c, 
the pair correlation functions
$P_{\nu}({\bf r}_0, {\bf r}) = |\Psi_{\nu}(({\bf r}_0, {\bf r})|^2$
where ${\bf r}_0$ is fixed at the center of the bottom dot.
$P_{\nu}({\bf r}_0, {\bf r})$ gives the probability of finding
the second electron at position {\bf r} given that the first electron
has been found at ${\bf r}_0$.  
For the ground state singlet $^1\Sigma_g^{(a)}$,
we see that at the small interdot separation $d$=4 nm, the probability
to find the second electron in the top or the bottom dot are
comparable, suggesting a molecular-like delocalized state. 
Accordingly, 
the wave function analysis reveals a dominant contribution from
the product of two delocalized molecular orbitals $\Phi_3$.
With increasing interdot separation, the electrons show 
correlation induced (i.e. the coupling between $\Phi_3$ and $\Phi_4$)
localization. At $d$=7 nm, the second electron
is almost entirely localized on the top dot as shown in 
Fig.~\ref{fig:corrfunc}c.    
A similar delocalized to localized transition  applies for the 
$^1\Sigma_g^{(b)}$ state, with the difference, that at large $d$ both
electrons are localized on the same (bottom) dot.
In contrast, the triplet states $^3\Sigma$ and the singlet $^1\Sigma_u$ show
localization at all interdot distances. 
However, for the triplet states the 
two electrons are localized on different dots while 
for $^1\Sigma_u$ both electrons are localized on the same dot.

\begin{figure}
\includegraphics[width=3.5in,angle=0]{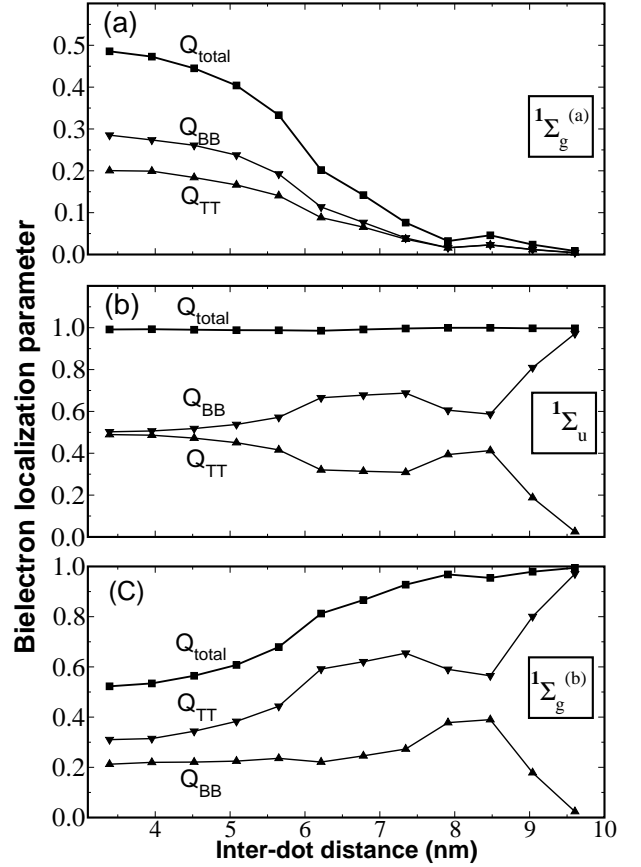}
\caption{
The probability of two electrons occupying the same dot 
[Eq.(\ref{eq:d_occ})]   for (a)
$^1\Sigma_g^{(a)}$, (b) $^1\Sigma_u$, (c) $^1\Sigma_g^{(b)}$
states. $Q_{\rm TT}$ ($Q_{\rm BB}$) is the probability of both electrons being
on the top (bottom) dot, while $Q_{\rm  total}$ gives the total probability.}
\label{fig:d_occ}
\end{figure}

To understand the (hidden) broken-symmetry (i.e. the wave function
localization)
in the many-particle CI
wave functions, and to study the degree of localization
quantitatively, we resort to the Wannier-like transformation of 
Eq.~(\ref{eq:wannier}). The CI matrix elements expressed
initially in the molecular basis $\psi_{i}$ [Eq.~(\ref{eq:wannier})] 
are transformed into
the dot-centered Wannier basis $\chi_{l,p}$ [Eq.~(\ref{eq:wannier})].
For example, $\{|\chi_{l,p}^{\sigma},\chi_{l',p'}^{\sigma'}\rangle\}$
denotes the
configuration where one electron is on the $l$-th orbital of the $p$ dot 
with spin
$\sigma$, and the other electron is on the $l'$-th orbital of the $p'$ dot 
with spin $\sigma'$.
The two electrons can be either both  on the top dots, or both on the bottom
dots, or one on the top and the other on the bottom dots. 
We can thus define a ``bielectron localization parameter'' 
$Q^{(\nu)}_{pp}$ as the probability of 
two electrons occupying the dot $p$= (T or B) at the same time in the
many-particle state $\nu$, 
\begin{equation}
Q^{(\nu)}_{pp}=\sum_{l\sigma,l'\sigma'} P_{\nu}(|\chi^{\sigma}_{l,p},
\chi^{\sigma'}_{l',p}\rangle) \, ,
\label{eq:d_occ}
\end{equation}
where $P_{\nu}(\mathcal{C})$ is the weight of the configuration $\mathcal{C}$ 
in the many-body wave functions of state $\nu$. The total probability
of two electrons being on the {\it same} dot is then 
$Q^{(\nu)}_{\rm tot}
=Q^{(\nu)}_{\rm TT}+Q^{(\nu)}_{\rm BB}$ for the $\nu$-th
state.   
Figure~\ref{fig:d_occ}, shows the bielectron localization parameter
$Q^{(\nu)}_{pp}$ of Eq.(\ref{eq:d_occ}) for
the many-particle states $\nu$= $^1\Sigma_g$,
$^1\Sigma_u$. We see that:
(i) For the ground state $^1\Sigma_g^{(a)}$ (Fig.~\ref{fig:d_occ}a), 
$Q_{\rm BB} > Q_{\rm TT}$ at all inter-dot separations 
(as is the case for the $^1\Sigma_u$ state), 
whereas for state $^1\Sigma_g^{(b)}$ (Fig.~\ref{fig:d_occ}c), 
we always have 
$Q_{\rm TT} > Q_{\rm BB}$, 
indicating symmetry breaking for these many-body
wave functions.
(ii) The ground state $^1\Sigma_g^{(a)}$ 
has a very small $Q_{pp}$ at large inter-dot separation ($d>$ 8
nm) , whereas for $^1\Sigma_g^{(b)}$, the probability of two
electrons being on the same dot is close to 1. 
At smaller $d$, $Q_{\rm tot}$ increases rapidly for 
$^1\Sigma_g^{(a)}$, while it decreases for   
$^1\Sigma_g^{(b)}$. 
(iii) Our calculations show that 
the coupling between $^3\Sigma$ and higher triplet 
configurations is negligible, thus
the calculated $Q_{pp}(^3\Sigma)$ are zero (not shown),
i.e. the electrons are on different dots, due to the Pauli principle.
However, in large electrostatic quantum dot
molecules\cite{ashoori92,johnson92,tarucha96} 
(with $J_{ee} > \Delta \epsilon_{e}$), where the lowest $^3\Sigma$ state 
is expected to mix with higher energy triplet configurations and 
acquire  non-zero double electron occupations,  $Q^{(\nu)}_{pp} > 0$.
(iv) $Q_{\rm tot}(^1\Sigma_u)$ is close to 1 for all
inter-dot distances (Fig.~\ref{fig:d_occ}b) and both electrons
are therefore occupying the same dots.
 
As we have noted in the introduction, quantum computing 
based on two spins in dot molecules requires that 
(i) $Q_{\rm tot}^{(\nu)} \ll$1 i.e. 
the electrons should be localized on different dots 
as much as possible. We now see this is satisfied for
$^3\Sigma_u$, but not for the
singlets $^1\Sigma_u$ and $^1\Sigma_g^{(b)}$. For $^1\Sigma_g^{(a)}$,
we find $Q_{\rm tot}^{(\nu)} \ll$1 only at large inter-dot separation $d$.
(ii) Two spins should have high entanglement. 
Since entanglement is maximized for pure singlet and triplet 
states without double-occupancy (i.e. $Q_{\rm tot}$=0), 
this also requires that $Q_{\rm tot}^{(\nu)} \ll$1.
(iii) The inter-dot separation should be such that 
significant singlet-triplet splitting exist.
Conditions (i) and (ii) require large inter-dot separation $d$, while (iii)
requires small $d$. 
Considering all the present results, the requirements for quantum computation 
are best met for the ground state $^1\Sigma_g^{(a)}$ at an inter-dot 
distance of 6 to 8 nm, where we have significant
$J_{S-T}$, and $Q_{\rm tot}$ is small. 

To conclude,  we investigated the wave function
asymmetry of both single-particle and many-particle wave functions 
in the quantum dot-molecules made of two identical,
vertically stacked InAs/GaAs self-assembled quantum dots.
For single-particle states, we find the asymmetry at
short interdot separations introduced by the inhomogeneous strains. 
For many-particle states, we find that
two electrons are always on different dots in the triplet state, 
while for the ground state singlet, the probability of
 ``bielectron localization''  is not zero and increases rapidly as the two
dots come close to each other. The ideal interdot separation where the two
electron spins can be used for quantum information is around 6 -8 nm. At this
distance the singlet-triplet splitting is large but the double 
electron occupation still low. 

This work was supported by US DOE-SC-BES-DMS, grant no. DEAC36-98-GO10337.



\newpage
\begin{thebibliography}{26}
\expandafter\ifx\csname natexlab\endcsname\relax\def\natexlab#1{#1}\fi
\expandafter\ifx\csname bibnamefont\endcsname\relax
  \def\bibnamefont#1{#1}\fi
\expandafter\ifx\csname bibfnamefont\endcsname\relax
  \def\bibfnamefont#1{#1}\fi
\expandafter\ifx\csname citenamefont\endcsname\relax
  \def\citenamefont#1{#1}\fi
\expandafter\ifx\csname url\endcsname\relax
  \def\url#1{\texttt{#1}}\fi
\expandafter\ifx\csname urlprefix\endcsname\relax\def\urlprefix{URL }\fi
\providecommand{\bibinfo}[2]{#2}
\providecommand{\eprint}[2][]{\url{#2}}

\bibitem[{\citenamefont{DiVincenzo}(1995)}]{divincenzo95}
\bibinfo{author}{\bibfnamefont{D.~P.} \bibnamefont{DiVincenzo}},
  \bibinfo{journal}{Science} \textbf{\bibinfo{volume}{270}},
  \bibinfo{pages}{255} (\bibinfo{year}{1995}).

\bibitem[{\citenamefont{Loss and DiVincenzo}(1998)}]{loss98}
\bibinfo{author}{\bibfnamefont{D.}~\bibnamefont{Loss}} \bibnamefont{and}
  \bibinfo{author}{\bibfnamefont{D.~P.} \bibnamefont{DiVincenzo}},
  \bibinfo{journal}{Phys. Rev. A} \textbf{\bibinfo{volume}{57}},
  \bibinfo{pages}{120} (\bibinfo{year}{1998}).

\bibitem[{\citenamefont{Schliemann et~al.}(2001)\citenamefont{Schliemann, Loss,
  and MacDonald}}]{schliemann01}
\bibinfo{author}{\bibfnamefont{J.}~\bibnamefont{Schliemann}},
  \bibinfo{author}{\bibfnamefont{D.}~\bibnamefont{Loss}}, \bibnamefont{and}
  \bibinfo{author}{\bibfnamefont{A.~H.} \bibnamefont{MacDonald}},
  \bibinfo{journal}{Phys. \ Rev. \ B} \textbf{\bibinfo{volume}{63}},
  \bibinfo{pages}{085311} (\bibinfo{year}{2001}).

\bibitem[{\citenamefont{Ashoori et~al.}(1992)\citenamefont{Ashoori, Stormer,
  Weiner, Pfeiffer, Pearton, Baldwin, and West}}]{ashoori92}
\bibinfo{author}{\bibfnamefont{R.~C.} \bibnamefont{Ashoori}},
  \bibinfo{author}{\bibfnamefont{H.~L.} \bibnamefont{Stormer}},
  \bibinfo{author}{\bibfnamefont{J.~S.} \bibnamefont{Weiner}},
  \bibinfo{author}{\bibfnamefont{L.~N.} \bibnamefont{Pfeiffer}},
  \bibinfo{author}{\bibfnamefont{S.~J.} \bibnamefont{Pearton}},
  \bibinfo{author}{\bibfnamefont{K.~W.} \bibnamefont{Baldwin}},
  \bibnamefont{and} \bibinfo{author}{\bibfnamefont{K.~W.} \bibnamefont{West}},
  \bibinfo{journal}{Phys. Rev. Lett.} \textbf{\bibinfo{volume}{68}},
  \bibinfo{pages}{3088} (\bibinfo{year}{1992}).

\bibitem[{\citenamefont{Johnson et~al.}(1992)\citenamefont{Johnson,
  Kouwenhoven, de~Jong, van~der Vaart, Harmans, and Foxon}}]{johnson92}
\bibinfo{author}{\bibfnamefont{A.~T.} \bibnamefont{Johnson}},
  \bibinfo{author}{\bibfnamefont{L.~P.} \bibnamefont{Kouwenhoven}},
  \bibinfo{author}{\bibfnamefont{W.}~\bibnamefont{de~Jong}},
  \bibinfo{author}{\bibfnamefont{N.~C.} \bibnamefont{van~der Vaart}},
  \bibinfo{author}{\bibfnamefont{C.~J. P.~M.} \bibnamefont{Harmans}},
  \bibnamefont{and} \bibinfo{author}{\bibfnamefont{C.~T.} \bibnamefont{Foxon}},
  \bibinfo{journal}{Phys.\ Rev.\ Lett.} \textbf{\bibinfo{volume}{69}},
  \bibinfo{pages}{1592} (\bibinfo{year}{1992}).

\bibitem[{\citenamefont{Tarucha et~al.}(1996)\citenamefont{Tarucha, Austing,
  Honda, van~der Hage, and Kouwenhoven}}]{tarucha96}
\bibinfo{author}{\bibfnamefont{S.}~\bibnamefont{Tarucha}},
  \bibinfo{author}{\bibfnamefont{D.~G.} \bibnamefont{Austing}},
  \bibinfo{author}{\bibfnamefont{T.}~\bibnamefont{Honda}},
  \bibinfo{author}{\bibfnamefont{R.~J.} \bibnamefont{van~der Hage}},
  \bibnamefont{and} \bibinfo{author}{\bibfnamefont{L.~P.}
  \bibnamefont{Kouwenhoven}}, \bibinfo{journal}{Phys.\ Rev.\ Lett.}
  \textbf{\bibinfo{volume}{77}}, \bibinfo{pages}{3613} (\bibinfo{year}{1996}).

\bibitem[{\citenamefont{Pi et~al.}(2001)\citenamefont{Pi, Emperador, Barranco,
  Garcias, Muraki, Tarucha, and Austing}}]{pi01}
\bibinfo{author}{\bibfnamefont{M.}~\bibnamefont{Pi}},
  \bibinfo{author}{\bibfnamefont{A.}~\bibnamefont{Emperador}},
  \bibinfo{author}{\bibfnamefont{M.}~\bibnamefont{Barranco}},
  \bibinfo{author}{\bibfnamefont{F.}~\bibnamefont{Garcias}},
  \bibinfo{author}{\bibfnamefont{K.}~\bibnamefont{Muraki}},
  \bibinfo{author}{\bibfnamefont{S.}~\bibnamefont{Tarucha}}, \bibnamefont{and}
  \bibinfo{author}{\bibfnamefont{D.~G.} \bibnamefont{Austing}},
  \bibinfo{journal}{Phys. Rev. Lett.} \textbf{\bibinfo{volume}{87}},
  \bibinfo{pages}{066801} (\bibinfo{year}{2001}).

\bibitem[{\citenamefont{Rontani et~al.}(2004)\citenamefont{Rontani, Amaha,
  Muraki, Manghi, Molinari, Tarucha, and Austing}}]{rontani04}
\bibinfo{author}{\bibfnamefont{M.}~\bibnamefont{Rontani}},
  \bibinfo{author}{\bibfnamefont{S.}~\bibnamefont{Amaha}},
  \bibinfo{author}{\bibfnamefont{K.}~\bibnamefont{Muraki}},
  \bibinfo{author}{\bibfnamefont{F.}~\bibnamefont{Manghi}},
  \bibinfo{author}{\bibfnamefont{E.}~\bibnamefont{Molinari}},
  \bibinfo{author}{\bibfnamefont{S.}~\bibnamefont{Tarucha}}, \bibnamefont{and}
  \bibinfo{author}{\bibfnamefont{D.~G.} \bibnamefont{Austing}},
  \bibinfo{journal}{Phys.\ Rev.\ B} \textbf{\bibinfo{volume}{69}},
  \bibinfo{pages}{085327} (\bibinfo{year}{2004}).

\bibitem[{\citenamefont{Fonseca et~al.}(1998)\citenamefont{Fonseca, Jimenez,
  Leburton, and Martin}}]{fonseca98}
\bibinfo{author}{\bibfnamefont{L.~R.~C.} \bibnamefont{Fonseca}},
  \bibinfo{author}{\bibfnamefont{J.~L.} \bibnamefont{Jimenez}},
  \bibinfo{author}{\bibfnamefont{J.~P.} \bibnamefont{Leburton}},
  \bibnamefont{and} \bibinfo{author}{\bibfnamefont{R.~M.}
  \bibnamefont{Martin}}, \bibinfo{journal}{Phys. Rev. B}
  \textbf{\bibinfo{volume}{57}}, \bibinfo{pages}{4017} (\bibinfo{year}{1998}).

\bibitem[{\citenamefont{Bednarek et~al.}(2003)\citenamefont{Bednarek, Szafran,
  Lis, and Adamowski}}]{bednarek03}
\bibinfo{author}{\bibfnamefont{S.}~\bibnamefont{Bednarek}},
  \bibinfo{author}{\bibfnamefont{B.}~\bibnamefont{Szafran}},
  \bibinfo{author}{\bibfnamefont{K.}~\bibnamefont{Lis}}, \bibnamefont{and}
  \bibinfo{author}{\bibfnamefont{J.}~\bibnamefont{Adamowski}},
  \bibinfo{journal}{Phys.\ Rev.\ B} \textbf{\bibinfo{volume}{68}},
  \bibinfo{pages}{155333} (\bibinfo{year}{2003}).

\bibitem[{\citenamefont{Hu and DasSarma}(2000)}]{hu00}
\bibinfo{author}{\bibfnamefont{X.}~\bibnamefont{Hu}} \bibnamefont{and}
  \bibinfo{author}{\bibfnamefont{S.}~\bibnamefont{DasSarma}},
  \bibinfo{journal}{Phys.\ Rev.\ A} \textbf{\bibinfo{volume}{61}},
  \bibinfo{pages}{62301} (\bibinfo{year}{2000}).

\bibitem[{\citenamefont{Rontani et~al.}(2001)\citenamefont{Rontani, Troiani,
  Hohenester, and Molinari}}]{rontani01}
\bibinfo{author}{\bibfnamefont{M.}~\bibnamefont{Rontani}},
  \bibinfo{author}{\bibfnamefont{F.}~\bibnamefont{Troiani}},
  \bibinfo{author}{\bibfnamefont{U.}~\bibnamefont{Hohenester}},
  \bibnamefont{and} \bibinfo{author}{\bibfnamefont{E.}~\bibnamefont{Molinari}},
  \bibinfo{journal}{Solid\ Sate\ Comm.} \textbf{\bibinfo{volume}{119}},
  \bibinfo{pages}{309} (\bibinfo{year}{2001}).

\bibitem[{\citenamefont{Yannouleas and Landman}(1999)}]{yannouleas99}
\bibinfo{author}{\bibfnamefont{C.}~\bibnamefont{Yannouleas}} \bibnamefont{and}
  \bibinfo{author}{\bibfnamefont{U.}~\bibnamefont{Landman}},
  \bibinfo{journal}{Phys. Rev. Lett.} \textbf{\bibinfo{volume}{82}},
  \bibinfo{pages}{5325} (\bibinfo{year}{1999}).

\bibitem[{\citenamefont{Stafford and Sarma}(1994)}]{stafford94}
\bibinfo{author}{\bibfnamefont{C.~A.} \bibnamefont{Stafford}} \bibnamefont{and}
  \bibinfo{author}{\bibfnamefont{S.~D.} \bibnamefont{Sarma}},
  \bibinfo{journal}{Phys. \ Rev. \ Lett.} \textbf{\bibinfo{volume}{72}},
  \bibinfo{pages}{3590} (\bibinfo{year}{1994}).

\bibitem[{\citenamefont{Bimberg et~al.}(1999)\citenamefont{Bimberg, Grundmann,
  and Ledentsov}}]{bimberg_book}
\bibinfo{author}{\bibfnamefont{D.}~\bibnamefont{Bimberg}},
  \bibinfo{author}{\bibfnamefont{M.}~\bibnamefont{Grundmann}},
  \bibnamefont{and} \bibinfo{author}{\bibfnamefont{N.~N.}
  \bibnamefont{Ledentsov}}, \emph{\bibinfo{title}{Quantum Dot
  Heterostructures}} (\bibinfo{publisher}{John Wiley \& Sons},
  \bibinfo{year}{1999}).

\bibitem[{\citenamefont{Ota et~al.}(2003)\citenamefont{Ota, Stopa, Rontani,
  Hatano, Yamada, Tarucha, Song, Nakata, Miyazawa, Ohshima et~al.}}]{ota03}
\bibinfo{author}{\bibfnamefont{T.}~\bibnamefont{Ota}},
  \bibinfo{author}{\bibfnamefont{M.}~\bibnamefont{Stopa}},
  \bibinfo{author}{\bibfnamefont{M.}~\bibnamefont{Rontani}},
  \bibinfo{author}{\bibfnamefont{T.}~\bibnamefont{Hatano}},
  \bibinfo{author}{\bibfnamefont{K.}~\bibnamefont{Yamada}},
  \bibinfo{author}{\bibfnamefont{S.}~\bibnamefont{Tarucha}},
  \bibinfo{author}{\bibfnamefont{H.}~\bibnamefont{Song}},
  \bibinfo{author}{\bibfnamefont{Y.}~\bibnamefont{Nakata}},
  \bibinfo{author}{\bibfnamefont{T.}~\bibnamefont{Miyazawa}},
  \bibinfo{author}{\bibfnamefont{T.}~\bibnamefont{Ohshima}},
  \bibnamefont{et~al.}, \bibinfo{journal}{Superlattices and Microstructures}
  \textbf{\bibinfo{volume}{34}}, \bibinfo{pages}{159} (\bibinfo{year}{2003}).

\bibitem[{\citenamefont{Ota et~al.}(2004)\citenamefont{Ota, Ono, Stopa, Hatano,
  Tarucha, Song, Nakata, Miyazawa, Ohshima, and Yokoyama}}]{ota04a}
\bibinfo{author}{\bibfnamefont{T.}~\bibnamefont{Ota}},
  \bibinfo{author}{\bibfnamefont{K.}~\bibnamefont{Ono}},
  \bibinfo{author}{\bibfnamefont{M.}~\bibnamefont{Stopa}},
  \bibinfo{author}{\bibfnamefont{T.}~\bibnamefont{Hatano}},
  \bibinfo{author}{\bibfnamefont{S.}~\bibnamefont{Tarucha}},
  \bibinfo{author}{\bibfnamefont{H.~Z.} \bibnamefont{Song}},
  \bibinfo{author}{\bibfnamefont{Y.}~\bibnamefont{Nakata}},
  \bibinfo{author}{\bibfnamefont{T.}~\bibnamefont{Miyazawa}},
  \bibinfo{author}{\bibfnamefont{T.}~\bibnamefont{Ohshima}}, \bibnamefont{and}
  \bibinfo{author}{\bibfnamefont{N.}~\bibnamefont{Yokoyama}},
  \bibinfo{journal}{Phys. \ Rev.\ Lett.} \textbf{\bibinfo{volume}{93}},
  \bibinfo{pages}{66801} (\bibinfo{year}{2004}).

\bibitem[{\citenamefont{Keating}(1966)}]{keating66}
\bibinfo{author}{\bibfnamefont{P.~N.} \bibnamefont{Keating}},
  \bibinfo{journal}{Phys. Rev.} \textbf{\bibinfo{volume}{145}},
  \bibinfo{pages}{637} (\bibinfo{year}{1966}).

\bibitem[{\citenamefont{Martins and Zunger}(1984)}]{martins84}
\bibinfo{author}{\bibfnamefont{J.~L.} \bibnamefont{Martins}} \bibnamefont{and}
  \bibinfo{author}{\bibfnamefont{A.}~\bibnamefont{Zunger}},
  \bibinfo{journal}{Phys.\ Rev.\ B} \textbf{\bibinfo{volume}{30}},
  \bibinfo{pages}{R6217} (\bibinfo{year}{1984}).

\bibitem[{\citenamefont{Williamson et~al.}(2000)\citenamefont{Williamson, Wang,
  and Zunger}}]{williamson00}
\bibinfo{author}{\bibfnamefont{A.~J.} \bibnamefont{Williamson}},
  \bibinfo{author}{\bibfnamefont{L.-W.} \bibnamefont{Wang}}, \bibnamefont{and}
  \bibinfo{author}{\bibfnamefont{A.}~\bibnamefont{Zunger}},
  \bibinfo{journal}{Phys.\ Rev.\ B} \textbf{\bibinfo{volume}{62}},
  \bibinfo{pages}{12963} (\bibinfo{year}{2000}).

\bibitem[{\citenamefont{Wang and Zunger}(1999)}]{wang99b}
\bibinfo{author}{\bibfnamefont{L.-W.} \bibnamefont{Wang}} \bibnamefont{and}
  \bibinfo{author}{\bibfnamefont{A.}~\bibnamefont{Zunger}},
  \bibinfo{journal}{Phys.\ Rev.\ B} \textbf{\bibinfo{volume}{59}},
  \bibinfo{pages}{15806} (\bibinfo{year}{1999}).

\bibitem[{\citenamefont{Bester and Zunger}(2005)}]{bester05a}
\bibinfo{author}{\bibfnamefont{G.}~\bibnamefont{Bester}} \bibnamefont{and}
  \bibinfo{author}{\bibfnamefont{A.}~\bibnamefont{Zunger}},
  \bibinfo{journal}{Phys. Rev. B} \textbf{\bibinfo{volume}{71}},
  \bibinfo{pages}{045318} (\bibinfo{year}{2005}).

\bibitem[{\citenamefont{Edmiston and Ruedenberg}(1963)}]{edmiston63}
\bibinfo{author}{\bibfnamefont{C.}~\bibnamefont{Edmiston}} \bibnamefont{and}
  \bibinfo{author}{\bibfnamefont{K.}~\bibnamefont{Ruedenberg}},
  \bibinfo{journal}{Rev. \ Mod. \ Phys.} \textbf{\bibinfo{volume}{35}},
  \bibinfo{pages}{457} (\bibinfo{year}{1963}).

\bibitem[{\citenamefont{He et~al.}()\citenamefont{He, Bester, and
  Zunger}}]{he05d}
\bibinfo{author}{\bibfnamefont{L.}~\bibnamefont{He}},
  \bibinfo{author}{\bibfnamefont{G.}~\bibnamefont{Bester}}, \bibnamefont{and}
  \bibinfo{author}{\bibfnamefont{A.}~\bibnamefont{Zunger}}, \bibinfo{note}{to
  be published.}

\bibitem[{\citenamefont{Franceschetti and Zunger}(1997)}]{Franceschetti97}
\bibinfo{author}{\bibfnamefont{A.}~\bibnamefont{Franceschetti}}
  \bibnamefont{and} \bibinfo{author}{\bibfnamefont{A.}~\bibnamefont{Zunger}},
  \bibinfo{journal}{Phys. \ Rev. \ Lett.} \textbf{\bibinfo{volume}{78}},
  \bibinfo{pages}{915} (\bibinfo{year}{1997}).

\bibitem[{\citenamefont{Franceschetti et~al.}(1999)\citenamefont{Franceschetti,
  Fu, Wang, and Zunger}}]{franceschetti99}
\bibinfo{author}{\bibfnamefont{A.}~\bibnamefont{Franceschetti}},
  \bibinfo{author}{\bibfnamefont{H.}~\bibnamefont{Fu}},
  \bibinfo{author}{\bibfnamefont{L.-W.} \bibnamefont{Wang}}, \bibnamefont{and}
  \bibinfo{author}{\bibfnamefont{A.}~\bibnamefont{Zunger}},
  \bibinfo{journal}{Phys.\ Rev.\ B} \textbf{\bibinfo{volume}{60}},
  \bibinfo{pages}{1819} (\bibinfo{year}{1999}).

\end{thebibliography}

\end {document}